\documentclass[aps,prd,onecolumn,preprintnumbers,groupedaddress,showpacs,
nofootinbib,amssymb]{revtex4}
\usepackage{graphicx}
\usepackage{epsfig}
\usepackage{bm}
\usepackage{amssymb}
\usepackage{float}
\usepackage{amsmath}
\usepackage{cancel}

\allowdisplaybreaks[4]
\newcommand{\e}{\mathrm{e}}

\begin{document}

\title{Novel cosmological and black hole solutions in Einstein and
higher-derivative gravity in two dimensions}

\author{Shin'ichi~Nojiri$^{1,2}$, 
Sergei~D.~Odintsov$^{3,4}$
}

\affiliation{ $^{1)}$ Department of Physics, Nagoya University,
Nagoya 464-8602, Japan \\
$^{2)}$ Kobayashi-Maskawa Institute for the Origin of Particles
and the Universe, Nagoya University, Nagoya 464-8602, Japan \\
$^{3)}$ ICREA, Passeig Luis Companys, 23, 08010 Barcelona, Spain\\
$^{4)}$ Institute of Space Sciences (IEEC-CSIC) C. Can Magrans
s/n, 08193 Barcelona, Spain\\ }

\begin{abstract}

We consider cosmological and black hole solutions in the Einstein and
higher-derivative gravity in two dimensions where the theory is formulated
first in $D$ dimensions. 
In the limit that $D$ tends to $2$ with simultaneous singular
rescaling of the scalar curvature coupling constant as $1/(D-2)$, we get 
the novel Einstein and higher-derivative gravity. 
Due to non-trivial contribution of scalar curvature which is topological 
invariant in exactly two dimensions to
gravitational equations in the two-dimensional limit one gets novel
cosmological
and black hole solutions.In particular, the de Sitter and 
radiation-dominated universe
and the Schwarzschild-de Sitter and Schwarzschild-anti-de Sitter 
black hole solutions are obtained and their properties are discussed.

\end{abstract}

\maketitle

\section{Introduction}

The study of dilatonic gravity at different dimensions attracts some attention
for a variety of reasons (for review, see Ref.~\cite{Nojiri:2000ja}).
In the first place, this theory is closely related with string theory, where it
appears as a sort of effective action.
A lot of activity is connected with the study of two-dimensional dilatonic
gravity, especially with its cosmology and black hole solutions.
Of course, at early studies of two-dimensional dilatonic gravity it was
repeated dozens times that the hope exists that through the study of easier 
two-dimensional models one can get some insights useful for the investigation 
of realistic four dimensional gravity.
However, the fundamental reason for the consideration of two-dimensional
dilatonic gravity is the fact that scalar curvature is topological invariant in 
two dimensions.
As a result two-dimensional analog of General Relativity with cosmological 
constant does not give any dynamics due to the absence of consistent 
gravitational equations.
The extra scalar degree of freedom suggests the way to overcome this difficulty
and leads to one of the forms of dilatonic gravity.

In this paper we show that non-trivial two-dimensional analog of General
Relativity may be formulated if we consider theory at $D$ dimensions first.
Then, the limit of $D$ tends to $2$ is considered while the coupling constant
of curvature invariant is rescaled by $1/(D-2)$ term.
In this way we get novel two-dimensional Einstein gravity which has
cosmological and black hole solutions.
Similar consideration for higher-derivative gravity
in two dimensions also opens the window for appearance of new cosmology 
and black hole solutions.
This is due to the fact that novel contributions to gravitational
equations appear from this scalar curvature invariant in the limit of two
dimensions.
In fact, this strategy follows from similar consideration of
Refs.~\cite{Glavan:2019inb,Fernandes:2020rpa} where the Einstein-Gauss-Bonnet
gravity was formulated in the limit from $D$ to $4$ dimensions with singular
rescaling of the Gauss-Bonnet coupling constant as $1/(D-4)$.

\section{Two-dimensional $R+\Lambda$ gravity}

\subsection{Exactly two-dimensional model}

We now consider the Einstein gravity in $D$ space-time dimensions
with a cosmological constant $\Lambda$
\begin{equation}
\label{2dE1}
S_\mathrm{Einstein}=\int d^D \sqrt{-g} \left( \frac{R}{\hat\alpha} + \Lambda
\right) \, .
\end{equation}
Here $\hat\alpha$ is a constant.
Then the Einstein equation is given by
\begin{equation}
\label{2dE2}
\frac{1}{\hat\alpha} \left( R_{\mu\nu} - \frac{1}{2} g_{\mu\nu} R \right)
= \frac{\Lambda}{2} g_{\mu\nu} \, ,
\end{equation}
which tells that the Ricci curvature $R_{\mu\nu}$ is covariantly constant
and therefore proportional to $g_{\mu\nu}$.
Multiplying Eq.~(\ref{2dE2}) with $g^{\mu\nu}$, we obtain
\begin{equation}
\label{2dE2B}
\frac{1}{\hat\alpha} \left( D-2 \right) R= \frac{D \Lambda}{2} \, .
\end{equation}
Thus, in two dimensions, $D=2$, there is no any solution, as is well-known.
As it was mentioned this is due to fact that scalar curvature is topological
invariant in two dimensions.

\subsection{The model in the limit of two dimensions}

We now redefine the parameter $\tilde\alpha$ as
\begin{equation}
\label{2dE3}
\tilde\alpha = \alpha \left(D-2\right) \, .
\end{equation}
Then Eq.~(\ref{2dE2B}) has the following form
\begin{equation}
\label{2dE4}
\frac{1}{\alpha} R= \frac{D \Lambda}{2} \, ,
\end{equation}
which has a solution in two dimensions,
\begin{equation}
\label{2dE5}
R = R_0 \equiv \frac{D \alpha \Lambda}{2} \, .
\end{equation}

When $R=R_0>0$, there is a solution describing the de Sitter space-time
\begin{equation}
\label{dSC1}
ds^2 = - dt^2 + \e^{\frac{t^2}{l^2}} \sum_{i=1}^{D-1} \left( dx^i \right)^2
\, , \quad R_0=\frac{D(D-1)}{l^2} \, .
\end{equation}
In the limit of two dimensions, that is, in the limit $D\to 2$,
the metric (\ref{dSC1}) has the following form,
\begin{equation}
\label{dSC2}
ds^2 = - dt^2 + \e^{\frac{t^2}{l^2}} dx^2 \, ,
\quad R_0=\frac{2}{l^2} \, .
\end{equation}
Hence, anyway there exists cosmological solution even in two space-time
dimensions.

On the other hand, as a static solution, we consider the solutions describing
black hole.\footnote{
It is interesting that in the $2D$ limit, the theory may be presented 
as scalar-tensor theory, see \cite{Mann:1992ar,Lu:2020iav}. }
If $R=R_0>0$, the Schwarzschild-de Sitter
space-time is an exact solution and if $R=R_0<0$, 
the Schwarzschild-anti-de Sitter space-time is an exact solution,
\begin{equation}
\label{DdSdS1}
ds^2 = - f(r) dt^2 + f(r)^{-1} dr^2 + r^2 d\Omega_{D-2}^2 \, , \quad
f(r) = 1- \frac{2M}{r^{D-3}} - \lambda r^2 \, , \quad
R_0=D(D-1)\lambda \, .
\end{equation}
Then in the limit $D\to2$, the metric in (\ref{DdSdS1}) has
the following form
\begin{equation}
\label{DdSdS5B}
ds^2 = - f(r) dt^2 + f(r)^{-1} dr^2 \, , \quad
f(r) = 1- 2Mr - \lambda r^2 \, , \quad
\lambda = \frac{R_0}{2} \, .
\end{equation}
When$R_0>0$, $\lambda \frac{1}{l^2}$ in (\ref{dSC1}) or (\ref{dSC2}).
Then there may appear horizons, where $f(r)=0$, at
\begin{equation}
\label{DdSdS6}
r=r_\pm \equiv - \frac{M\pm \sqrt{M^2 + \lambda}}{\lambda} \, .
\end{equation}
In order that there exists real and positive solution, we should require
\begin{equation}
\label{DdSdS7}
0>\lambda>-M^2\, .
\end{equation}
Then $r=r_-$ corresponds to the inner horizon and $r=r_+$ to the outer
horizon as for the charged black hole.
The extremal limit, where the radii of the two horizons coincide with each
other, is given by $M^2 = - \lambda$.
Thus, we obtained a new type of black hole solution in two-dimensional General
Relativity.

We now rewrite $f(r)$ in the following form,
\begin{equation}
\label{DdSdS8}
f(r)=\frac{\left( r - r_+ \right) \left( r - r_- \right)}{r_+ r_-} \, .
\end{equation}
When $r\sim r_\pm$, by defining $\delta r$ by $r=r_\pm \pm \delta r$, we find
\begin{equation}
\label{DdSdS9}
f(r)=\frac{\left( r_+ - r_- \right)\delta r}{r_+ r_-} \, .
\end{equation}
Then by the Wick rotating the time coordinate $t\to i\tau$, the metric
(\ref{DdSdS5B}) has the following form,
\begin{equation}
\label{DdSdS10}
ds^2=\frac{\left( r_+ - r_- \right)\delta r}{r_+ r_-} d\tau^2
+ \frac{r_+ r_-}{\left( r_+ - r_- \right)\delta r} dr^2 \, .
\end{equation}
By defining a new coordinate $\rho$ as
\begin{equation}
\label{DdSdS11}
d\rho = \pm dr \sqrt{ \frac{r_+ r_-}{\left( r_+ - r_- \right)\delta r}} \quad
\mbox{that is,} \quad
\rho = \pm 2 \sqrt{ \frac{ r_+ r_- \delta r}{r_+ - r_-}} \quad
\mbox{or} \quad \delta r= \frac{r_+ - r_-}{4 r_+ r_-}\rho^2 \, ,
\end{equation}
the metric (\ref{DdSdS10}) can be rewritten as
\begin{equation}
\label{DdSdS12}
ds^2=\frac{\left( r_+ - r_- \right)^2}{ 4 r_+^2 r_-^2} \rho^2d\tau^2
+ d\rho^2 \, .
\end{equation}
In order to avoid the conical singularity in the Wick-rotated Euclidean space,
we require the periodicity,
\begin{equation}
\label{DdSdS13}
\frac{r_+ - r_-}{ 2r_+ r_-} \tau
\sim \frac{r_+ - r_- }{ 2r_+ r_-} \tau + 2\pi \, ,
\end{equation}
which shows that black hole has the temperature $T$ as
\begin{equation}
\label{DdSdS14}
T=\frac{r_+ - r_-}{ 4\pi r_+ r_-}
= \frac{\sqrt{M^2 + \lambda}}{\pi} \, .
\end{equation}
There occurs also the Hawking radiation.
Eq.~(\ref{DdSdS14}) shows that the temperature $T$ vanishes in the
extremal limit $r_+=r_-$ or $M^2=-\lambda$.
Note that it is difficult to define the thermodynamical quantities like entropy
in two dimensions because the area of the black hole horizon naively vanishes.
The area of $D-2$ dimensional unit sphere is given by
\begin{equation}
\label{DdSdS15}
S_{D-2} = \frac{\pi^{\frac{D-1}{2}}}{\Gamma\left( \frac{D-1}{2} \right)} \, .
\end{equation}
Here $\Gamma$ is the gamma function.
When $D=2$, the expression (\ref{DdSdS15}) gives $S_0=2$, which corresponds to
the two points whose center is the origin on the straight line.
We may assume that the entropy is proportional to $S_{D-1}/G$
where $G$ is Newton's gravitational constant.
The action (\ref{2dE1}) tells that $\tilde\alpha$ can be regarded to be
proportional to
$G$ and therefore the entropy should be proportional to $S_{D-1}/\tilde\alpha$.
If we keep $\alpha$ in (\ref{2dE3}) to be finite, the entropy diverges in the
limit of $D\to 2$ and therefore the entropy is ill-defined.

\section{$R+R^2$ gravity}

Let us consider $F(R)$ gravity action without matter in
$D$ space-time dimensions (for general introduction, see
Refs.~\cite{Nojiri:2010wj, Nojiri:2017ncd,Capozziello:2011et})
\begin{equation}
\label{JGRG7}
S_{F(R)}= \frac{1}{2\kappa^2} \int d^D x \sqrt{-g} F(R) \, .
\end{equation}
Then the equation of motion for the $F(R)$ gravity is given by
\begin{equation}
\label{JGRG13}
0=\frac{1}{2}g_{\mu\nu} F(R) - R_{\mu\nu} F'(R) - g_{\mu\nu} \Box F'(R)
+ \nabla_\mu \nabla_\nu F'(R) \, .
\end{equation}
If we assume that the Ricci curvature $R_{\mu\nu}$ is covariantly constant and
the scalar curvature $R$ is constant, Eq.~(\ref{JGRG13}) reduces to
\begin{equation}
\label{JGRG13B}
0=\frac{1}{2}g_{\mu\nu} F(R) - R_{\mu\nu} F'(R) \, .
\end{equation}
Multiplying $g^{\mu\nu}$, we obtain an algebraic equation for $R$,
\begin{equation}
\label{JGRG13C}
0=\frac{D}{2} F(R) - R F'(R) \, .
\end{equation}
If Eq.~(\ref{JGRG13C}) has a positive solution, $R=R_0>0$,
the Schwarzschild-de Sitter space-time is an exact solution and if $R=R_0<0$,
the Schwarzschild-anti-de Sitter
space-time is an exact solution as in (\ref{DdSdS1}).

\subsection{Exactly two-dimensional model}

We now consider the following model
\begin{equation}
\label{DdSdS2B}
F(R)=\frac{R}{\tilde\alpha} + \beta R^2 \, .
\end{equation}
It was shown long ago \cite{Muta:1992xw} that
higher-derivatives terms for such model and its generalizations give
constant curvature solution in the way similar to the Jackiw-Teitelboim gravity
\cite{Teitelboim:1983ux,Jackiw:1984je}.
Then Eq.~(\ref{JGRG13C}) has the following form,
\begin{equation}
\label{DdSdS3B}
0= \frac{\left(D-2\right) R}{2\tilde\alpha} + \frac{D-4} \beta R^2 \, ,
\end{equation}
In two dimensions, $D=2$, the solution is given by
\begin{equation}
\label{DdSdS3B2}
R=0 \, .
\end{equation}
If $R=0$, Eq.~(\ref{JGRG13}) with (\ref{DdSdS2B}) requires
\begin{equation}
\label{DdSdS2BB}
R_{\mu\nu}=0\, .
\end{equation}

First we consider the cosmological solution by assuming the FRW space-time
\begin{equation}
\label{FRW}
ds^2 = - dt^2 + a(t)^2 \sum_{i=1}^{D-1} \left( dx^i \right)^2 \, .
\end{equation}
In the $D$ dimensional FRW space-time, the curvatures have the following
expression,
\begin{equation}
\label{curvatures}
R_{tt} = - \left( D-1 \right) \left( \dot H + H^2 \right) \, , \quad
R_{ij} = a^2 \left( \dot H + \left( D -1 \right) H^2 \right) \delta_{ij} \, ,
\quad
R = \left( 2\left( D - 1 \right) \dot H + D \left( D-1 \right) H^2 \right) \, ,
\end{equation}
Here $H=\frac{\dot a}{a}$.
Especially in two dimensions, $D=2$, the expressions (\ref{curvatures}) reduce
to
\begin{equation}
\label{2dcurvatures}
R_{tt} = - \left( \dot H + H^2 \right) \, , \quad
R_{ij} = a^2 \left( \dot H + H^2 \right) \delta_{ij} \, , \quad
R = 2 \left(\dot H + H^2 \right) \, ,
\end{equation}
Then Eqs.~(\ref{DdSdS2BB}) is satisfied if
\begin{equation}
\label{2dFRFRW}
0 = \dot H + H^2 \, .
\end{equation}
whose solution is given by
\begin{equation}
\label{2dFRFRW2}
H = \frac{1}{t-t_0} \, , \quad
a=a_0 \left( t-t_0 \right) \, ,
\end{equation}
with constants $t_0$ and $a_0$.
The solution (\ref{2dFRFRW2}) corresponds to the radiation dominated
universe in four dimensions but as a model in two dimensions, this solution
gives a new type of cosmology.

We should note that there is a static solution (\ref{DdSdS5B}) with $\lambda=0$
even if $R=0$ in (\ref{DdSdS3B2}),
\begin{equation}
\label{DdSdS5B2}
ds^2 = - f(r) dt^2 + f(r)^{-1} dr^2 \, , \quad
f(r) = 1- 2Mr \, ,
\end{equation}
which might look similar to the Schwarzschild solution and a
horizon at $r=r_0 \equiv \frac{1}{2M}$ but the solution does not describe
any black hole solution because the space-time signature is given by
$(+,-,+,+)$ when $r>r_0$ although the signature is given by
$(-,+,+,+)$ for the black hole solution.
Hence the horizon at $r=r_0 \equiv \frac{1}{2M}$ is not the black hole
horizon but the cosmological horizon.
Note that there is no solution in the Einstein gravity in
two dimensions as mentioned after Eq.~(\ref{2dE2}) but there is anyway a
solution (\ref{DdSdS3B2}), which is because the $F(R)$ gravity can be rewritten
in the form of the scalar-tensor theory, which gives a solution.
Therefore the solution (\ref{DdSdS5B2}) could be new type of static solution.

We can consider more complicated models of the sort introduced in
Ref.~\cite{Muta:1992xw} like
\begin{equation}
\label{muta}
F(R) = \sum_i a_i R^{n_i} \, ,
\end{equation}
where $a_i$ are coupling constants and powers $n_i$ are some numbers.
Then, such models have constant curvature solutions which give the 
de Sitter, Schwarzschild-de Sitter, or Schwarzschild-anti-de Sitter space-time.

\subsection{The model in the limit of two dimensions}

Instead of the model (\ref{DdSdS2B}) by redefining the parameter $\tilde\alpha$
as in (\ref{2dE3}) we consider the following theory
\begin{equation}
\label{DdSdS2}
F(R)=\frac{R}{\alpha\left(D-2\right)} + \beta R^2 \, .
\end{equation}
Then Eq.~(\ref{JGRG13C}) has the following form,
\begin{equation}
\label{DdSdS3}
0= \frac{R}{2\alpha} + \frac{D-4} \beta R^2 \, ,
\end{equation}
and we find non-trivial solution even in two dimensions,
\begin{equation}
\label{DdSdS4}
R_0= 0,\ - \frac{1}{2(D-4)\alpha\beta} \, .
\end{equation}
In the limit $D\to2$, $R_0$ is finite.
Hence, if $R_0=0$, we have the cosmological solution as in
(\ref{2dFRFRW2})
\begin{equation}
\label{2dFRFRW2B}
H = \frac{1}{t-t_0} \, , \quad
a=a_0 \left( t-t_0 \right) \, ,
\end{equation}
This cosmology corresponds to two-dimensional radiation-dominated universe.
If $R_0>0$, the cosmological solution describing the de Sitter
space-time,
\begin{equation}
\label{dSC2B}
ds^2 = - dt^2 + \e^{\frac{t^2}{l^2}} dx^2 \, ,
\quad \frac{1}{l^2} = \frac{R_0}{2} = \frac{1}{8\alpha\beta} \, ,
\end{equation}
and we also obtain the static space-time in (\ref{DdSdS5B})
\begin{equation}
\label{DdSdS1BB}
ds^2 = - f(r) dt^2 + f(r)^{-1} dr^2 + r^2 d\Omega_{D-2}^2 \, , \quad
f(r) = 1- \frac{2M}{r^{D-3}} - \lambda r^2 \, ,
\end{equation}
with
\begin{equation}
\label{DdSdS5B0BB}
\lambda =0,\ \frac{1}{8\alpha\beta} \, .
\end{equation}
Then as in the previous section, this solution may give a new type of black
hole space-time in two dimensions.
When $\lambda\neq 0$, there may appear two horizons as in (\ref{DdSdS6}),
\begin{equation}
\label{DdSdS6BBB}
r=r_\pm \equiv - \frac{M\pm \sqrt{M^2 + \lambda}}{\lambda} \, .
\end{equation}
This shows that theory (\ref{DdSdS2}) has a solution describing the black hole
in the limit of $D\to 2$, whose situation is different from the model
(\ref{DdSdS2B}).
The black hole has a temperature $T$ as in (\ref{DdSdS14}),
\begin{equation}
\label{DdSdS14BBB}
T = \frac{\sqrt{M^2 + \lambda}}{\pi} \, ,
\end{equation}
and therefore there occurs the Hawking radiation.
Hence, even in case of higher-derivative gravity in this limit from $D$ to $2$
with
singular rescaling of coupling constant we get novel cosmological and black
hole solutions.
It is easy to see that this scheme may be easily generalized for
more complicated versions of higher-derivative gravity \cite{Muta:1992xw}.
Again, due to non-trivial contribution from $R$ term the cosmological and
the black hole dynamics will be changed.

\section{Conclusion and Discussion}

We formulated the novel Einstein and $R^2$ gravity in two dimensions as the limit
from $D$-dimensional theory to two-dimensional theory with simultaneous singular
rescaling of gravitational coupling constant as $1/(D-2)$. As a result even the 
two-dimensional Einstein gravity becomes non-trivial theory which contains 
the de Sitter cosmology and the Schwarzschild-de Sitter or 
Schwarzschild-anti-de Sitter black hole solutions. 
This occurs due to non-trivial contribution of scalar curvature term in such limit to
gravitational equations. 
For $R^2$ gravity we also got novel black hole solutions
as well as the de Sitter universe and radiation-dominated universe cosmologies.

It would be interesting to further study this approach in related
two-dimensional dilatonic gravities.
Let us consider
Callan-Giddings-Harvey-Strominger (CGHS) model
\cite{Callan:1992rs} (see also \cite{Nojiri:1992ug}).
The action is given by
\begin{equation}
\label{CGHS1}
S_\mathrm{CGHS} = \frac{1}{2\pi} \int d^2 x \sqrt{-g} \left[
\e^{-2\phi} \left( R + 4 \left( \nabla \phi \right)^2 + 4\lambda^2 \right)
 - \frac{1}{2} \left( \nabla f \right)^2 \right] \, .
\end{equation}
Here $\phi$ and $f$ are dilaton and matter fields, respectively and
$\lambda^2$ is a cosmological constant.
This model has a solution describing the black hole in exactly two dimensions.
By using the light-cone coordinates, $x^\pm = t \pm x$,
\begin{equation}
\label{CGHS2}
ds^2 = - \left( \frac{M}{\lambda} - \lambda^2 x^+ x^- \right)dx^+ dx^- \, .
\end{equation}
Then there appears one horizon at $x^+ x^-=\frac{M}{\lambda^2}$,
which is different from the models in this paper where there appear two
horizons, (\ref{DdSdS6}) and (\ref{DdSdS6BBB}).
In \cite{Callan:1992rs}, the black hole formation was discussed by using the
exact shock wave solution and the Hawking radiation and back reaction of the
metric is analyzed by using the trace-anomaly.
We should note that the Hawking temperature is independent of the black hole
mass and given by $T=\frac{1}{2\pi}$, which is also different from the
temperature in the models of this paper, (\ref{DdSdS14}) and (\ref{DdSdS14BBB}).

It might be interesting to extend the CGHS model (or 
the Jackiw-Teitelboim gravity with quantum
matter fields, see for instance, Ref.~\cite{Moitra:2019xoj}) 
by adding scalar curvature term with
gravitational coupling constant and considering such theory in the same limit
from $D$ to $2$ dimensions. 
Then again scalar curvature term gives non-trivial
contribution to gravitational equations what can significantly change its
dynamics. However, one should also account for quantum effects in such limit
(say, conformal anomaly). Hence, such generalization requests careful
investigation.

\begin{acknowledgments} 
This work is partially supported  by MEXT KAKENHI Grant-in-Aid for 
Scientific Research on Innovative Areas ``Cosmic Acceleration'' No. 15H05890 (S.N.) 
and the JSPS Grant-in-Aid for Scientific Research (C) No. 18K03615 (S.N.), 
and by MINECO (Spain), FIS2016-76363-P (S.D.O). 
\end{acknowledgments}


\begin{thebibliography}{99}

\bibitem{Nojiri:2000ja}
S.~Nojiri and S.~D.~Odintsov,
Int.\ J.\ Mod.\ Phys.\ A {\bf 16} (2001) 1015
doi:10.1142/S0217751X01002968
[hep-th/0009202].


\bibitem{Glavan:2019inb}
D.~Glavan and C.~Lin,
Phys.\ Rev.\ Lett.\ {\bf 124} (2020) no.8, 081301
doi:10.1103/PhysRevLett.124.081301
[arXiv:1905.03601 [gr-qc]].

\bibitem{Fernandes:2020rpa}
P.~G.~S.~Fernandes,
arXiv:2003.05491 [gr-qc].

\bibitem{Mann:1992ar}
R.~B.~Mann and S.~Ross,
Class.\ Quant.\ Grav.\  \textbf{10} (1993), 1405-1408
doi:10.1088/0264-9381/10/7/015
[arXiv:gr-qc/9208004 [gr-qc]].

\bibitem{Lu:2020iav}
H.~Lu and Y.~Pang,
arXiv:2003.11552 [gr-qc].

\bibitem{Teitelboim:1983ux}
C.~Teitelboim,
Phys.\ Lett.\ {\bf 126B} (1983) 41.
doi:10.1016/0370-2693(83)90012-6

\bibitem{Jackiw:1984je}
R.~Jackiw,
Nucl.\ Phys.\ B {\bf 252} (1985) 343.
doi:10.1016/0550-3213(85)90448-1

\bibitem{Callan:1992rs}
C.~G.~Callan, Jr., S.~B.~Giddings, J.~A.~Harvey and A.~Strominger,
Phys.\ Rev.\ D {\bf 45} (1992) no.4, R1005
doi:10.1103/PhysRevD.45.R1005
[hep-th/9111056].

\bibitem{Nojiri:1992ug}
S.~Nojiri and I.~Oda,
Nucl.\ Phys.\ B {\bf 406} (1993) 499
doi:10.1016/0550-3213(93)90179-S
[hep-th/9207077].

\bibitem{Nojiri:2010wj}
S.~Nojiri and S.~D.~Odintsov,
Phys.\ Rept.\ \textbf{505} (2011), 59-144
doi:10.1016/j.physrep.2011.04.001
[arXiv:1011.0544 [gr-qc]].

\bibitem{Nojiri:2017ncd}
S.~Nojiri, S.~D.~Odintsov and V.~K.~Oikonomou,
Phys.\ Rept.\ {\bf 692} (2017) 1
doi:10.1016/j.physrep.2017.06.001
[arXiv:1705.11098 [gr-qc]].

\bibitem{Capozziello:2011et}
S.~Capozziello and M.~De Laurentis,
Phys.\ Rept.\ {\bf 509} (2011) 167
doi:10.1016/j.physrep.2011.09.003
[arXiv:1108.6266 [gr-qc]].

\bibitem{Muta:1992xw}
T.~Muta and S.~Odintsov,
Prog.\ Theor.\ Phys.\ \textbf{90} (1993) 247-255
doi:10.1143/PTP.90.247

\bibitem{Moitra:2019xoj}
U.~Moitra, S.~K.~Sake, S.~P.~Trivedi and V.~Vishal,
[arXiv:1908.08523 [hep-th]].

\end{thebibliography}
\end{document}